\documentclass[12pt]{article}
\usepackage{amssymb,amsmath,epsfig}
\usepackage{graphicx}
\usepackage{subfig}
\usepackage{cite}
 \allowdisplaybreaks
\begin{document}
\title{\bf Energy Conditions in Modified $f(G)$ Gravity}

\author{Kazuharu Bamba$^1$ \thanks{bamba@sss.fukushima-u.ac.jp}, M. Ilyas $^2$
\thanks{ilyas\_mia@yahoo.com}, M. Z. Bhatti$^3$ \thanks{mzaeem.math@pu.edu.pk} and
Z. Yousaf$^3$ \thanks{zeeshan.math@pu.edu.pk}\\
$^1$ Division of Human Support System,\\ Faculty of Symbiotic
Systems Science,\\ Fukushima University, Fukushima 960-1296, Japan\\
$^{2}$Centre for High Energy Physics, University of the Punjab\\
$^{3}$ Department of Mathematics, University of the Punjab,\\
Quaid-i-Azam Campus, Lahore-54590, Pakistan}

\date{}

\maketitle
\begin{abstract}
In this paper, we have considered flat
Friedmann-Lema\^{i}tre-Robertson-Walker metric in the framework of
perfect fluid models and modified $f(G)$ gravity (where $G$ is the
Gauss Bonnet invariant). Particularly, we have considered particular
realistic $f(G)$ configurations that could be used to cure
finite-time future singularities arising in the late-time cosmic
accelerating epochs. We have then developed the viability bounds of
these models induced by weak and null energy conditions, by using
the recent estimated numerical figures of the deceleration, Hubble,
snap and jerk parameters.
\end{abstract}
{\bf Keywords:} Relativistic fluids; Modified gravity; Stability\\
{\bf PACS:} 04.50.Kd; 04.20.-q; 98.80.Jk; 98.80.-k\\
{\bf Report number:} FU-PCG-20

\section{Introduction}

Several interesting outcomes stem from observations of Supernovae
Type Ia, cosmic microwave background radiation, etc. \cite{ya1} have
produced a revolution in the field of relativistic astrophysics and
cosmology. This has created a new alluring platform for research.
These ingredients have revealed that current expansion of universe
is accelerating. The observational data came from, e.g., the Planck
satellite~\cite{ya2, Planck:2015xua, Ade:2015lrj}, the BICEP2
experiment~\cite{Ade:2014xna, Ade:2015tva, Array:2015xqh}, and the
Wilkinson Microwave anisotropy probe (WMAP)~\cite{Komatsu:2010fb,
Hinshaw:2012aka}, have illustrated that the energy fractions of the
baryonic and dark matter (DM) are 5$\%$ and 27$\%$, respectively,
while that of dark energy (DE) is only 68$\%$. The concept of
modified gravity theories (MGTs) obtained by replacing the Ricci
scalar in the standard Einstein-Hilbert (EH) action with some generic
functions of the Ricci scalar $f(R)$ or the combinations of the
scalar and tensorial curvature invariants have been introduced by
many relativistic astrophysicists. This approach has now been
referred as a standard terminology whose formulations could be
considered as a viable guide to explore reason of the cosmic
accelerated expansion (for further reviews on dark energy and
modified gravity, see, for
instance,~\cite{R1,R2,R3,R4,R5,R6,R7,R8,R9,R10,3,4,5,6,8}). The
first consistent outcomes of accelerating universe from $f(R)$
gravity was suggested by Nojiri and Odintsov \cite{no1}. There has
been an interested results found on the exploration of dark source
terms on the dynamical evolution of stellar systems in
Einstein-$\Lambda$ \cite{zs1}, $f(R)$ \cite{z1fr}, $f(R,T)$
\cite{z2frt} ($T$ is the trace of energy momentum tensor) and
$f(R,T,R_{\mu\nu}T^{\mu\nu})$ gravity \cite{z3frtrmn}.

Among MGTs, available in the literature, the one is Gauss Bonnet
(GB) gravity which has received great attraction
\cite{9,10,11,12,13,14,15,16,17,18,19} and is named as $f(G)$
gravity, where $G = R - 4{R_{\mu \nu }}{R^{\mu \nu }} + {R_{\mu \nu
\alpha \beta }}{R^{\mu \nu \alpha \beta }}$ is a topological
invariant in 4 dimensions of spacetime. The equation of motion for
this gravity is required to be coupled with some scalar field or $f(G)$ must be
some arbitrary function of $G$. This MGT could help out in the study
of inflationary era, transition of acceleration from deceleration
regimes, passing tests induced by solar system experiments and
crossing phantom divide line for different viable $f(G)$ models
\cite{11,12}. It is also seen that the GB gravity is less
constrained than $f(R)$ gravity \cite{13}. , The $f(G)$ gravity also
provide an efficient platform to study various cosmic issues as an
alternate to DE \cite{14}. The $f(G)$ gravity could also be very
helpful for the study of finite time future singularities as well as
the universe acceleration during late time epochs \cite{15,16}.
Similarly, the cosmic acceleration followed by matter era could also
be explained by means of some viable models in $f(G)$ gravity
\cite{13,14}. Different consistent $f(G)$ models were proposed in order
to pass certain solar system constraints \cite{13,14} which are
discussed in \cite{17} and additional bounds on $f(G)$ models may
arise from the analysis of energy conditions (ECs) \cite{18,19,20}. Nojiri \emph{et al.} \cite{bin4} have
discussed some fundamental cosmic issues, like inflation, late-time acceleration, bouncing cosmology and
claimed that some modified theories of gravity, like
$f(R),~f(G)$ and $f(\mathcal{T})$ theories (where $\mathcal{T}$ is the torsion scalar) could be used as a viable mathematical tool
for analyzing the clear picture of our universe.

The ECs are the basic ingredients for the deep understanding of the
singularity theorem as like the theorem of black-hole
thermodynamics. Hawking-Penrose singularity theorem imposed the
importance of the weak energy (WE) and strong energy (SE)
conditions, while the black hole second law of thermodynamics
signifies null energy (NE) condition. The well-known Raychaudhuri
equations could be considered to discuss the viability of various
forms of ECs \cite{21,22,23}. Some of the literature review of
ECs were discussed by using the classical ECs of
general relativity (GR) like the phantom fields potential \cite{24},
the history of expanding universe \cite{25,26,27,28,29,30} and the
pattern movement of deceleration parameters \cite{31,32}. The
various expression for ECs are derived in $f(R)$ gravity \cite{33}
and using these formalism and techniques, some authors have pointed out
some issues (cosmological) in $f(R)$ gravity \cite{34,35,36}. The general
formalism for ECs are derived in $f(G)$ gravity by Garc\'{i}a \emph{et al.} \cite{37}.
Nojiri \emph{et al.} \cite{15} presented
some specific realistic and viable $f(G)$ models by analyzing the
dynamical behavior of WEC. Garc\'{i}a \emph{et al.} \cite{gar1} have
explored some viable $f(G)$ models and checked their viability
epochs by exploring ECs. Sadeghi \emph{et al.} \cite{sad1} have
explored some $f(G)$ garvity models that could obey WEC and SEC in
an era where late-time de-Sitter solution was stable. Banijamali
\emph{et al.} \cite{bani1} analyzed the distribution of WEC for a
class of consistent $f(G)$ models and claimed that power law model
of the type $f(G)=\epsilon G^n$ would satisfy WEC on setting
$\epsilon<0$.

In this paper, we have used some of the approximate values of the
jerk, deceleration, Hubble as well as snap model parameters, we then
apply certain limits from $f(G)$ gravity ECs on the model building
variables which were suggested in paper \cite{13}. We showed by
different plots that these models in $f(G)$ gravity can satisfy the
WEC and SEC in a specific region which is necessary for exploring
the stability of a late time de-Sitter solutions. This work is
formatted in a manner that the coming section consists of brief
introduction to $f(G)$ field equations as well as modified version
of ECs. In section \textbf{3}, we shall consider some viable $f(G)$
models in order to explore the viability epochs of ECs. The
conclusions and results are summarized in the last section.

\section{Field equation}

This section is devoted to illustrate the extended version of GB gravity with
its equations of motion as well as ECs. For $f(G)$ gravity, the usual EH action is modified as follows
\begin{equation}\label{action}
S = \int {{d^4}x\sqrt { - g} \left[ {\frac{R}{2} + f(G)} \right] + {S_M}\left( {{g^{\mu \nu }},\psi } \right)},
\end{equation}
where ${\kappa ^2} = 8\pi G \equiv 1$, $R,~f,~{S_M}({{g^{\mu \nu }},\psi})$ are the Ricci
scalar, arbitrary function of GB invariant and the matter action, respectively. The GB invariant quantity is
\begin{equation}
G = R - 4{R_{\mu \nu }}{R^{\mu \nu }} + {R_{\mu \nu \alpha \beta }}{R^{\mu \nu \alpha \beta }},
\end{equation}
where $R_{\mu \nu}$ is the Ricci tensor and ${R_{\mu \nu \alpha \beta }}$ is the Riemannian tensor. Upon varying the above action
with respect to $g_{\mu \nu }$, we get the modified field equations for $f(G)$ gravity as
\begin{equation}\label{field equation}
{R_{\mu \nu }} - \frac{1}{2}R{g_{\mu \nu }} = T_{\mu \nu
}^{\textrm{eff}},
\end{equation}
where $T_{\mu \nu }^{\textrm{eff}}$ is dubbed as effective energy
momentum tensor with its expression as follows
\begin{align}\nonumber
T_{\mu \nu }^{\textrm{eff}}&= {\kappa ^2}{T_{\mu \nu }} - 8\left[
{{R_{\mu \rho \nu \sigma }}} \right. + {R_{\mu \nu }}{g_{\mu \nu }}
- {R_{\mu \nu }}{g_{\mu \nu }} - {R_{\mu \nu }}{g_{\mu \nu
}}+{R_{\mu \nu }}{g_{\mu \nu }}\\\label{field} &+
\frac{1}{2}({g_{\mu \nu }}{g_{\mu \nu }}
 - {g_{\mu \nu }}\left. {{g_{\mu \nu }})} \right]{\nabla ^\rho }{\nabla ^\sigma }f_G
 + \left( {G f_G - f} \right){g_{\mu \nu }},
\end{align}
where subscript $G$ defines the derivation of the corresponding term
with the GB term, while ${T_{\mu \nu }}$ is the usual stress energy
momentum tensor. We model our system with the following well-known
line element of Friedmann-Lema\^{i}tre-Robertson-Walker (FLRW)
universe
\begin{equation}
d{s^2} =  - d{t^2} + {a^2}(t)(d{x^2} + d{y^2} + d{z^2}),
\end{equation}
in which $a(t)$ is the scale factor. We assume that this line element is
filled with an ideal matter content whose energy momentum tensor is
$$T_{\mu \nu }=diag(\rho(r),-p(r),-p(r),-p(r)).$$
In this context, the $f(G)$ field equations (\ref{field equation}) turns out to be
\begin{equation}
{\rho^{\textrm{eff}}} = \frac{3}{2}{H^2},\quad {p^{\textrm{eff}}} =
- \left( {2H' + 3{H^2}} \right)
\end{equation}
where prime symbolizes for temporal derivations, ${\rho ^{\textrm{eff}}}$ and
${p ^{\textrm{eff}}}$ are effective energy density and the pressure gradient,
respectively. For FLRW universe filled with perfect fluid, the
expressions for effective energy density and the pressure component
become
\begin{align}\nonumber
{\rho ^{\textrm{eff}}} &= \rho  + \frac{1}{2}\left[- f(G) +
24{H^2}f'(G)({H^2} + H') - 576{H^4}{f^{\prime \prime }}(G)(4{H^2}H'
+ 2{{H'}^2}\right.\\\label{z1} &\left.+ H{H^{\prime \prime }})
\right],\\\nonumber {p^{\textrm{eff}}}&=p+\frac{1}{2}\left[ f(G) -
24{H^2}f'(G)({H^2} + H')
 + 8H\{576{H^3}(4{H^2}H' + 2{{H'}^2} + H
 \right.\\\nonumber
 &\times\left. {H^{\prime \prime }})^2f^{(3)}(G)+ 24H{f^{\prime \prime }}(G)(8{H^4}H'
 +6H'^3 + 6{H^3}H^{\prime \prime } + 8HH'{H^{\prime \prime }}+H^2\right.\\\label{z2}
 &\times\left.(24H'^2 + H^{(3)}))\} \right]
\end{align}
The GB and the Ricci invariants for the flat FLRW spacetime are
found as follows
\begin{align}\nonumber
G&= 24{H^2}\left( {H' + {H^2}} \right), \quad R= 6\left( {H' + 2{H^2}} \right).
\end{align}

\section{Energy Conditions}

In different physical scenario, the basic and fundamental tools for
the study of black holes, wormholes (WHs) etc, are the ECs. The
breaching of these constraints could be fruitful to analyze the
stability of WHs. The situation of exploring ECs in MGTs is quite
different because the field equations differ from the Einstein
equations. The ECs in GR are derived by relating $R_{\mu\nu}$ with
usual energy momentum tensor. In MGTs, such a relation is not
straightforward. One must know how to relate $R_{\mu\nu}$ with the
effective forms of energy momentum tensor which will eventually give
rise to the corresponding ECs. These ECs are the outcomes of
Raychaudhuri's equation for the expansion nature. In MGTs (having
effective energy density and pressure), the NEC and WEC are defined
as follows
\begin{align}
&\textrm{NEC}\Leftrightarrow {\rho^{\textrm{eff}}} +
{p^{\textrm{eff}}} \ge 0,\\\nonumber &\textrm{WEC}\Leftrightarrow
{\rho^{\textrm{eff}}} \ge 0 \text{ and } {\rho^{\textrm{eff}}} +
{p^{\textrm{eff}}} \ge 0,
\end{align}
while the SEC and the dominant energy condition (DEC) provide
\begin{align}
&\textrm{SEC}\Leftrightarrow {\rho^{\textrm{eff}}+
3{p^{\textrm{eff}}}} \ge 0 \text{ and } {\rho^{\textrm{eff}}} +
{p^{\textrm{eff}}} \ge 0,\\\nonumber &\textrm{DEC} \Leftrightarrow
{\rho^{\textrm{eff}}} \ge 0 \text{ and } {\rho^{\textrm{eff}}} \pm
{p^{\textrm{eff}}} \ge 0.
\end{align}
We see that ECs would impose some constraints on the parameters
involved in the building of $f(G)$ models \cite{37}. It has been
clear that the derivative of position four vector is referred as
four velocity and its double derivative is termed as four
acceleration. Further, its third and fourth derivatives give jerk
and snap parameters, respectively. The Hubble parameter for FLRW
metric filled with an ideal matter is found as follows
\begin{align}\label{z3}
H=\frac{\dot{a}}{a},
\end{align}
while deceleration $q$, jerk $j$ and snap $s$ parameters turn out to
be
\begin{align}\label{z4}
q =-\frac{1}{{{H^2}}}\frac{{a''}}{a},\quad j =
\frac{1}{{{H^3}}}\frac{{a'''}}{a}, \quad s =
\frac{1}{{{H^4}}}\frac{{{a''''}}}{a}.
\end{align}
By means of these parameters, the derivatives of Hubble parameters
become
\begin{align}\label{z5}
H'=-H{^2}(q+1),~~{H^{(3)}} = H{^4}(- 2j- 5q+s-3),~~ H''=H{^3}(j + 3q
+ 2).
\end{align}
Using Eqs.(\ref{z3})-(\ref{z5}), Eqs.(\ref{z1}) and (\ref{z2}) can
be recasted as
\begin{align}\nonumber
{\rho^{\textrm{eff}}} + {p^{\textrm{eff}}} &= \rho  + p +
\frac{1}{2}\left[ {192{H^8}} \right.((3 - 14q - 24{q^2} - 6{q^3}
 - j\left( {7 + 8q} \right) + s)\\\label{effropp}
 &\times{f^{\prime \prime }}(G) + 24{H^4}{\left( {j + q\left( {3 + 2q} \right)} \right)^2}\left.
 {{f^{\left( 3 \right)}}(G)} \right],\\\label{effro}
{\rho^{\textrm{eff}}} &= \rho  + \frac{1}{2}\left[ { - f(G)} \right.
- 24({H^4}qf'(G) + 24{H^8}\left( {j + q\left( {3 + 2q} \right)}
\right)\left. {{f^{\prime \prime }}(G)} \right].
\end{align}
It is worth noticing that above equations have been expressed by
taking into account arbitrary function of $G$.

\section{Specific Models}

In this section, we shall check the influences of some $f(G)$
models, with vacuum (i.e. $\rho=p=0$) background, on the
formulations and behavior of the ECs. In the following calculation,
we would use the following specific numerical values of Hubble,
deceleration, snap and jerk parameters \cite{38}
$$H=0.718,~~q=-0.64,~~j=1.02,~~s=-0.39.$$

The following subsections would allow us to set up various
configurations of FLRW models controlled by few particular $f(G)$
models.

\subsection{Model 1}

\begin{figure} \centering
\epsfig{file=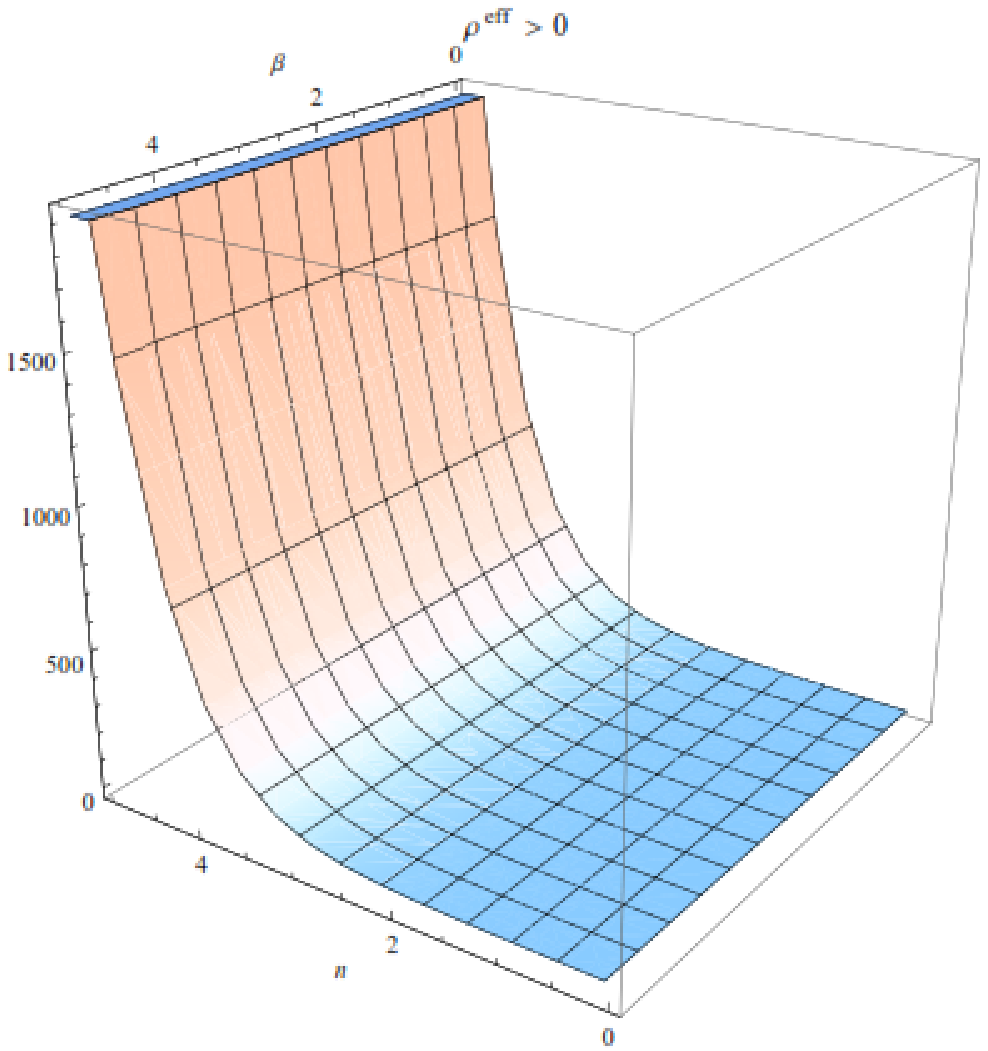,width=.48\linewidth}
\epsfig{file=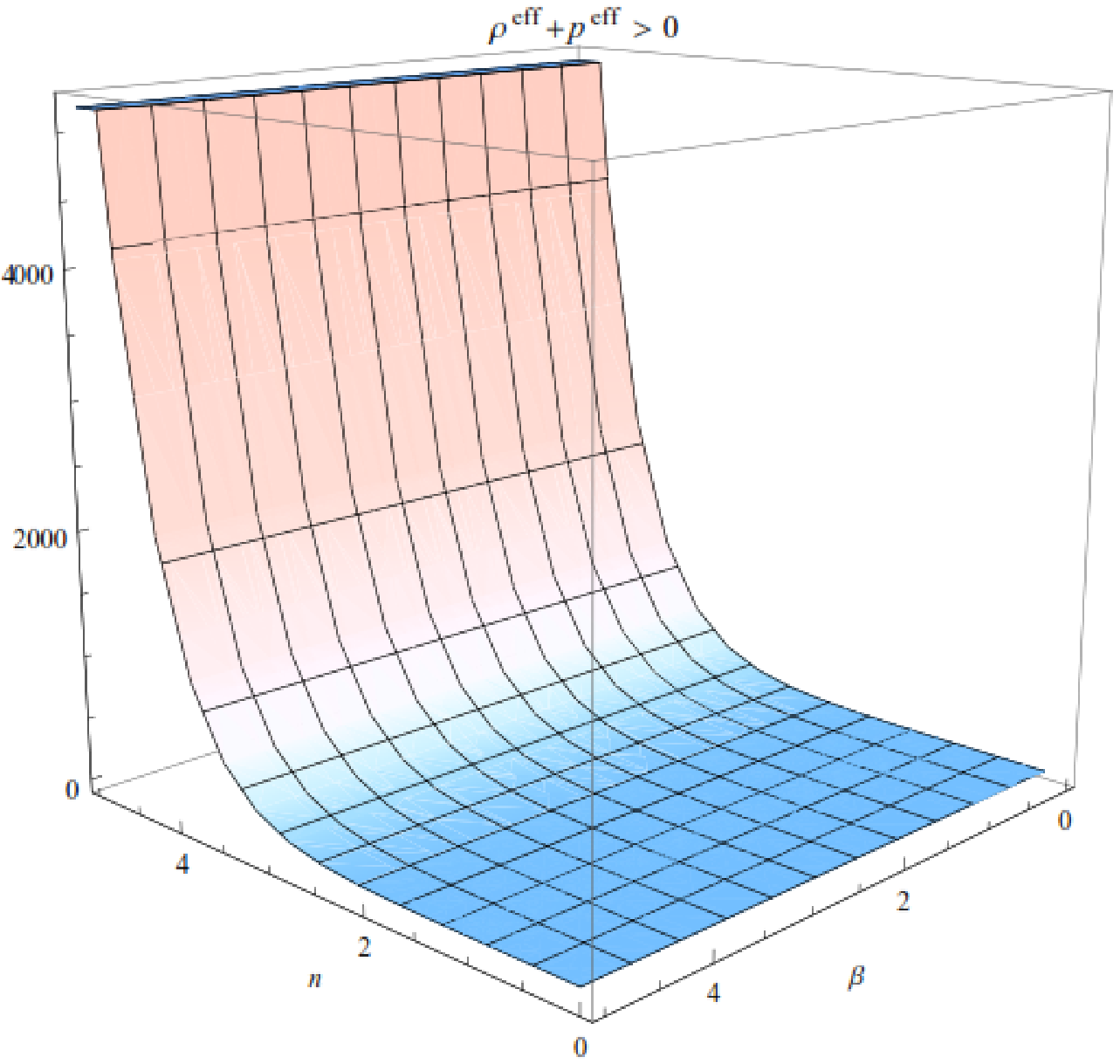,width=.48\linewidth} \caption{WEC plots for
$f(G)$ model given in Eq.(\ref{model1}). Here, left and right plots are
showing the behaviors of $\rho^{\textrm{eff}}$ and $\rho^{\textrm{eff}}+
p^{\textrm{eff}}$ with respect to $\beta$ and $n$ for
$\alpha=0.3$, respectively.} \label{1f1}
\end{figure}

First, we assume a model containing combinations of power law and
logarithmic $f(G)$ corrections \cite{39}
\begin{equation}\label{model1}
f(G) = \alpha {G^n} + \beta G\log[G],
\end{equation}
where $\alpha,~n$ and $\beta$ are constants. The dynamics presented
by this model is found to be in agreement with the data
presented by the same cosmographic parameters \cite{40}. This upon
substituting in Eq.(\ref{effro}), we obtain the effective energy
density as
\begin{align}\nonumber
{\rho ^{\textrm{eff}}}&= \frac{1}{{{q^2}}}(24\beta {H^4}q(j + q(3 + q)) -
{24^n}(-1+ n)\alpha( - {H^4}q)^n(nj + q(3n\\\label{1ro} &+( - 1 +
2n)q)))\geq 0,
\end{align}
while the sum of effective pressure energy density can be obtained,
after using Eqs.(\ref{effropp}) and (\ref{model1}), as follows
\begin{align}\nonumber
{p^{\textrm{eff}}+\rho ^{\textrm{eff}}} &= \frac{1}{{3{q^3}}}(3q(24\beta {H^4}q -
{24^n}( - 1 + n)n\alpha {( - {H^4}q)^n})\left( {j + q\left( {3 + 2q}
\right)} \right)\\\nonumber
 &- \left( {24\beta {H^4}q + {{24}^n}n\left( {2 - 3n + {n^2}} \right)
 \alpha {{( - {H^4}q)}^n}} \right)(j{\left( {j + q\left( {3 + 2q} \right)}
 \right)^2}\\\nonumber
 &+ q\left( {24\beta {H^4}q - {{24}^n}\left( { - 1 + n} \right)n\alpha {{( - {H^4}q)}^n}} \right)
 (-3 + 5q + 18{q^2} + 6{q^3}\\\label{1ropp}
 &+ j\left( {4 + 8q} \right) - s))\geq 0.
\end{align}
To get an exact solution from the above two inequalities (\ref{1ro})
and (\ref{1ropp}), for the parameters $\alpha$, $\beta$ and $n$, is
a quite hard task. In order to achieve this goal, we would consider
specific value of $\alpha=0.3$ and plot $\rho^{\textrm{eff}}$ and
$\rho^{\textrm{eff}}+ p^{\textrm{eff}}$ as a function of $\beta$ and
$n$ as shown in fig.\ref{1f1}. One can see the validity of WECs from
the fig.\ref{1f1}.

\begin{figure} \centering
\epsfig{file=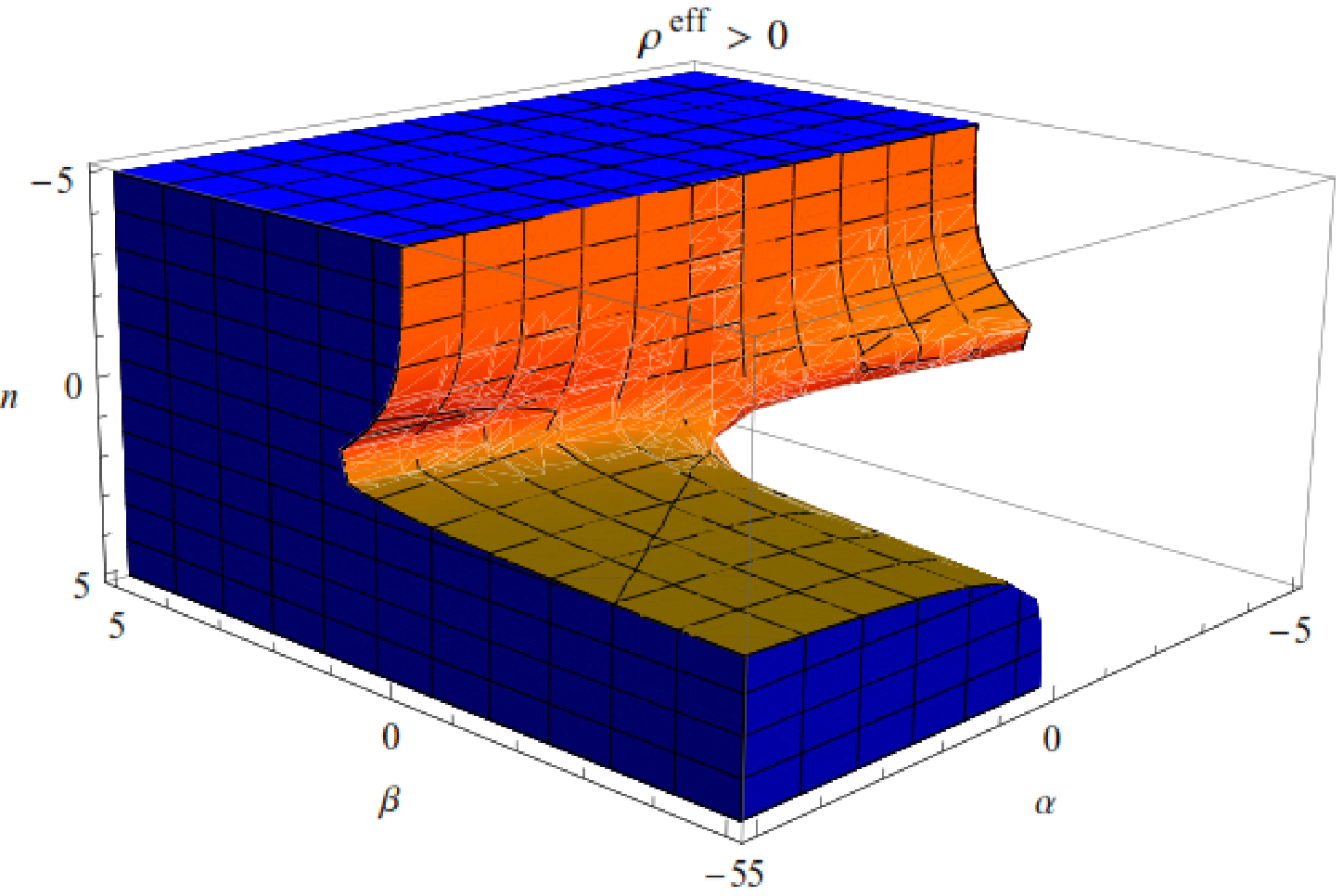,width=.48\linewidth}
\epsfig{file=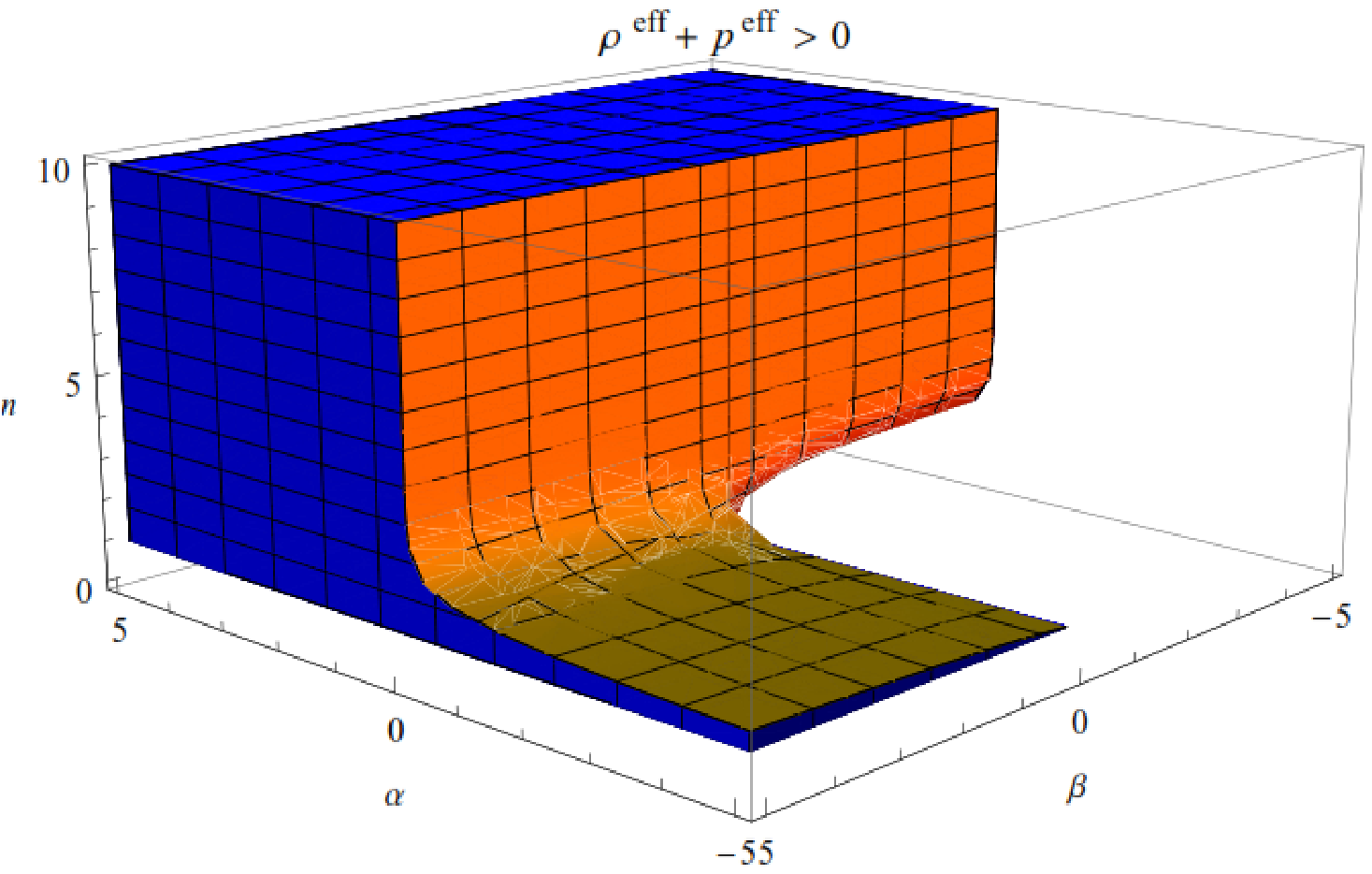,width=.48\linewidth} \caption{WEC plots for
$f(G)$ model mentioned in Eq.(\ref{model1}). here, the left and right plots
are representing the regions where $\rho_{eff}>0$ and $\rho_{eff}+ p_{eff}>0$
with respect to $\alpha$, $\beta$ and
$n$, respectively. We see that the WEC is satisfied for the considered range of
parameters.} \label{1f2}
\end{figure}

Now we will discuss the constraints which is required for the validity of WEC,
i.e., for $\rho^{\textrm{eff}}\geq 0$. The validity of WEC
is guaranteed if \\
(1) $\alpha>0$ with $n>1$ and for all values of $\beta$.\\
(2) $\alpha>0$ require $n<-1$ and $\beta>0$.\\
We took the range of parameters as $\alpha,\beta,n\in(-5,5)$.
The validity region is shown in the left plot of fig.\ref{1f2}.
Similarly, if we take $n>1$ then $\rho^{\textrm{eff}}+ p^{\textrm{eff}}\geq 0$ imposed
the constraints on the parameters as\\
(1) $\alpha>0$ with for all value of $\beta$.\\
(2) for very small $n$ with for all $\alpha$ and  $\beta>0$.\\
The plotted region in this context are shown in the right diagram of
fig.\ref{1f2}. We have concluded that on taking all the parameters
($n,~\alpha,~\beta$) to be positive in $f(G) = \alpha {G^n} + \beta
G\log[G]$ model, the violation of WEC can be avoided.

\subsection{Model 2}

\begin{figure} \centering
\epsfig{file=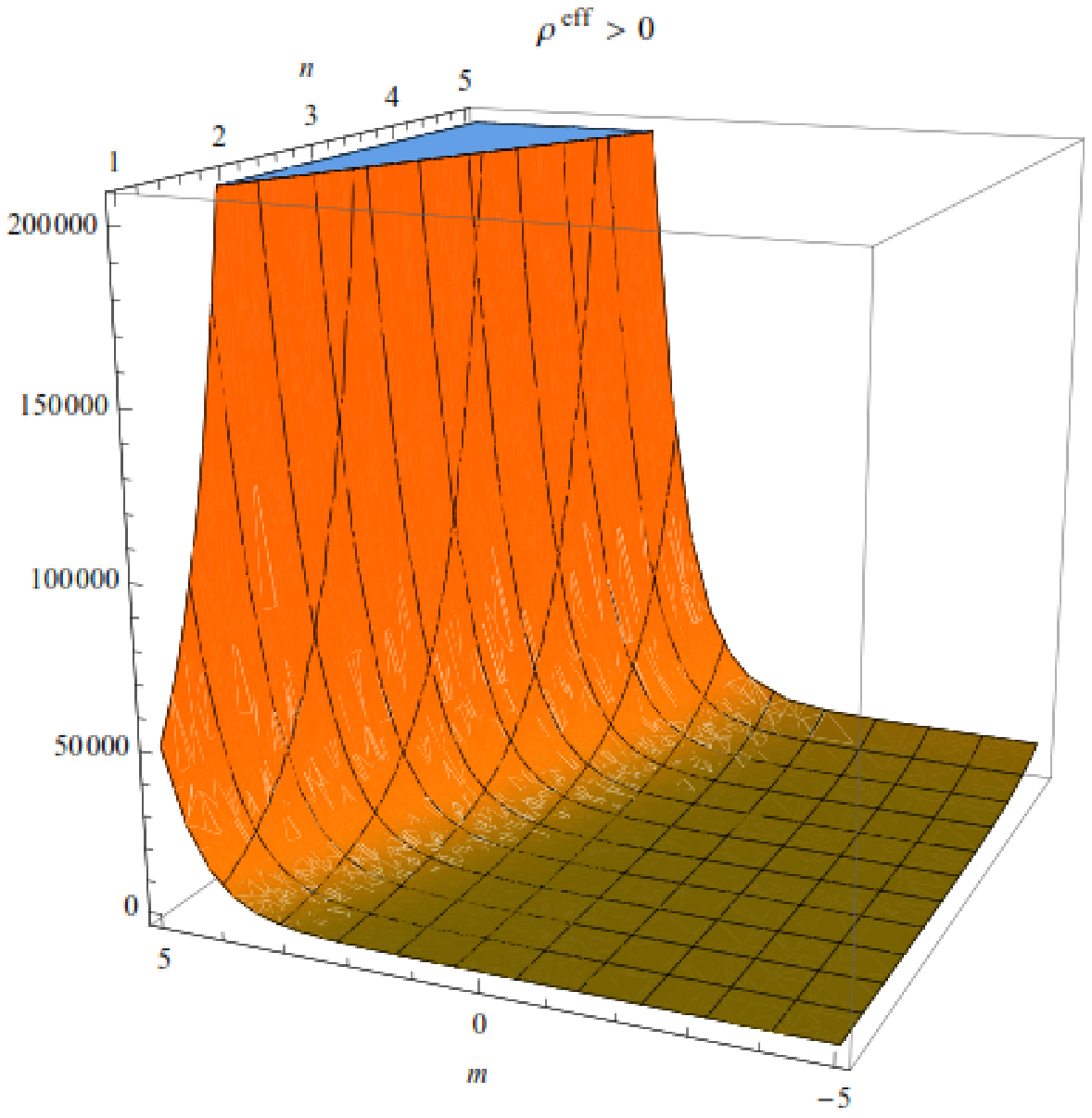,width=.48\linewidth}
\epsfig{file=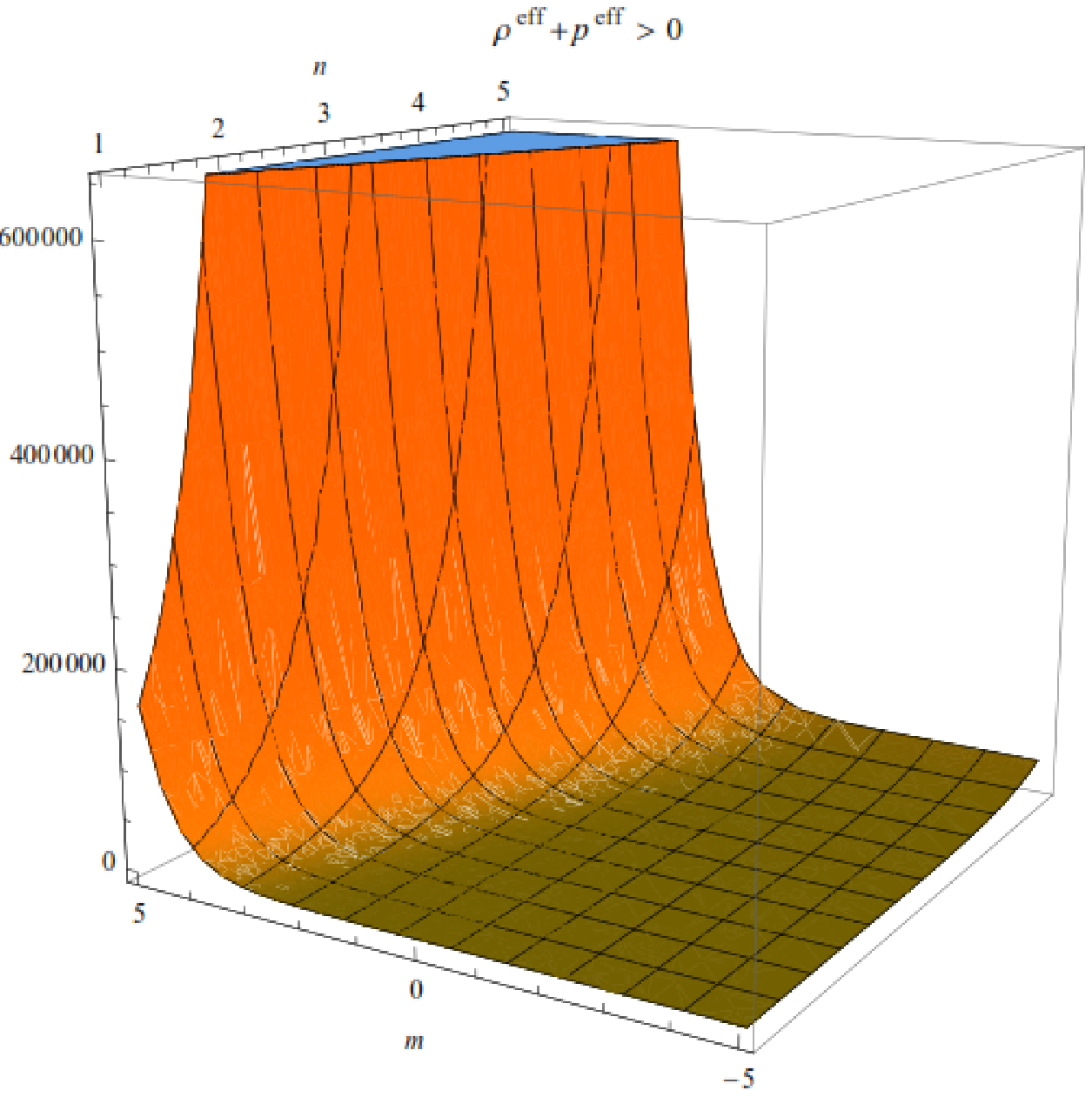,width=.48\linewidth} \caption{Plots of WEC
for $f(G)$ model given in Eq.(\ref{model2}), In this figure, the left and right plots show
the distributions of $\rho^{\textrm{eff}}$ and $\rho^{\textrm{eff}}+ p^{\textrm{eff}}$
with respect to $m$ and $n$ with $\alpha=1$, $\beta=1$, respectively.} \label{2f1}
\end{figure}

\begin{figure} \centering
\epsfig{file=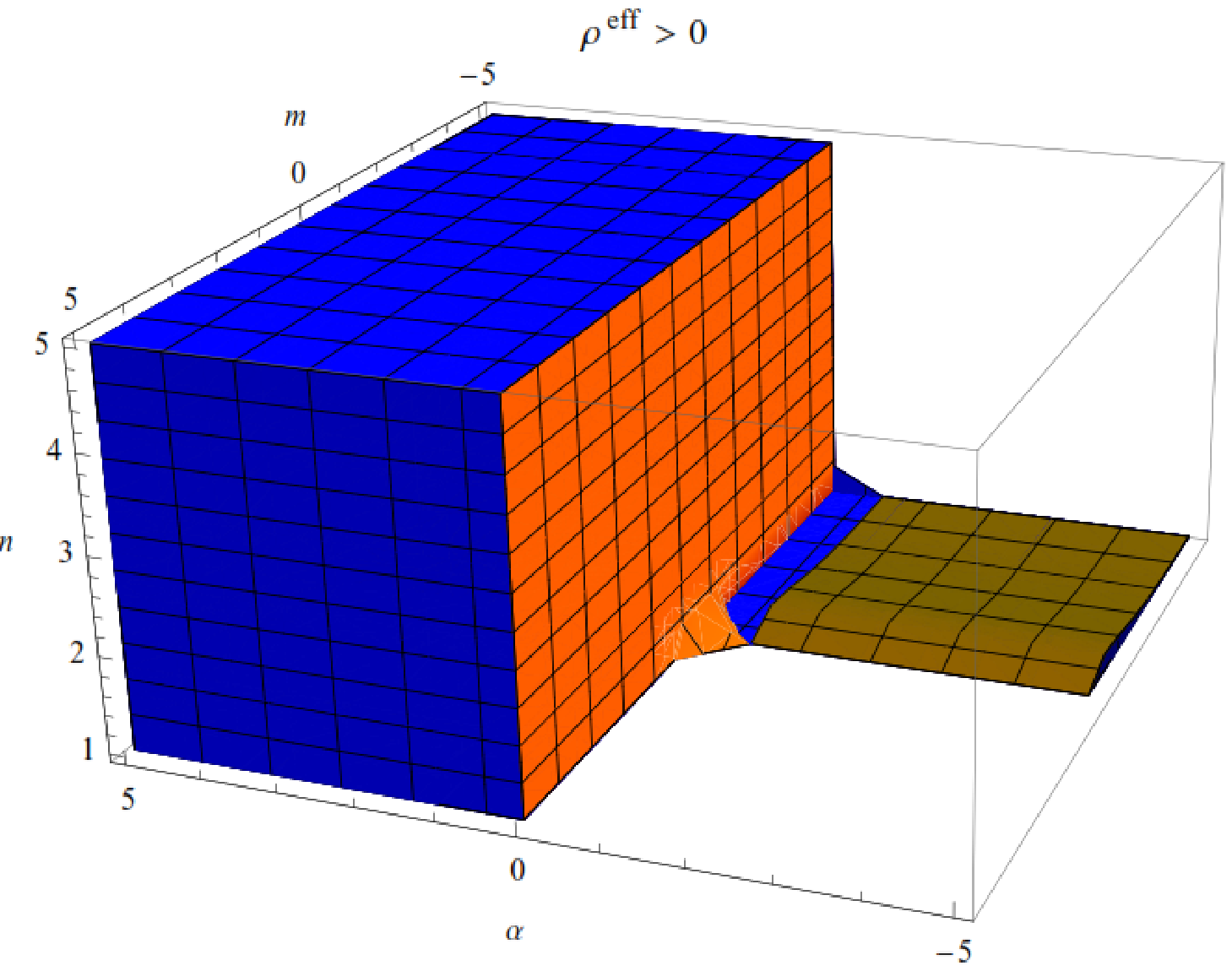,width=.48\linewidth}
\epsfig{file=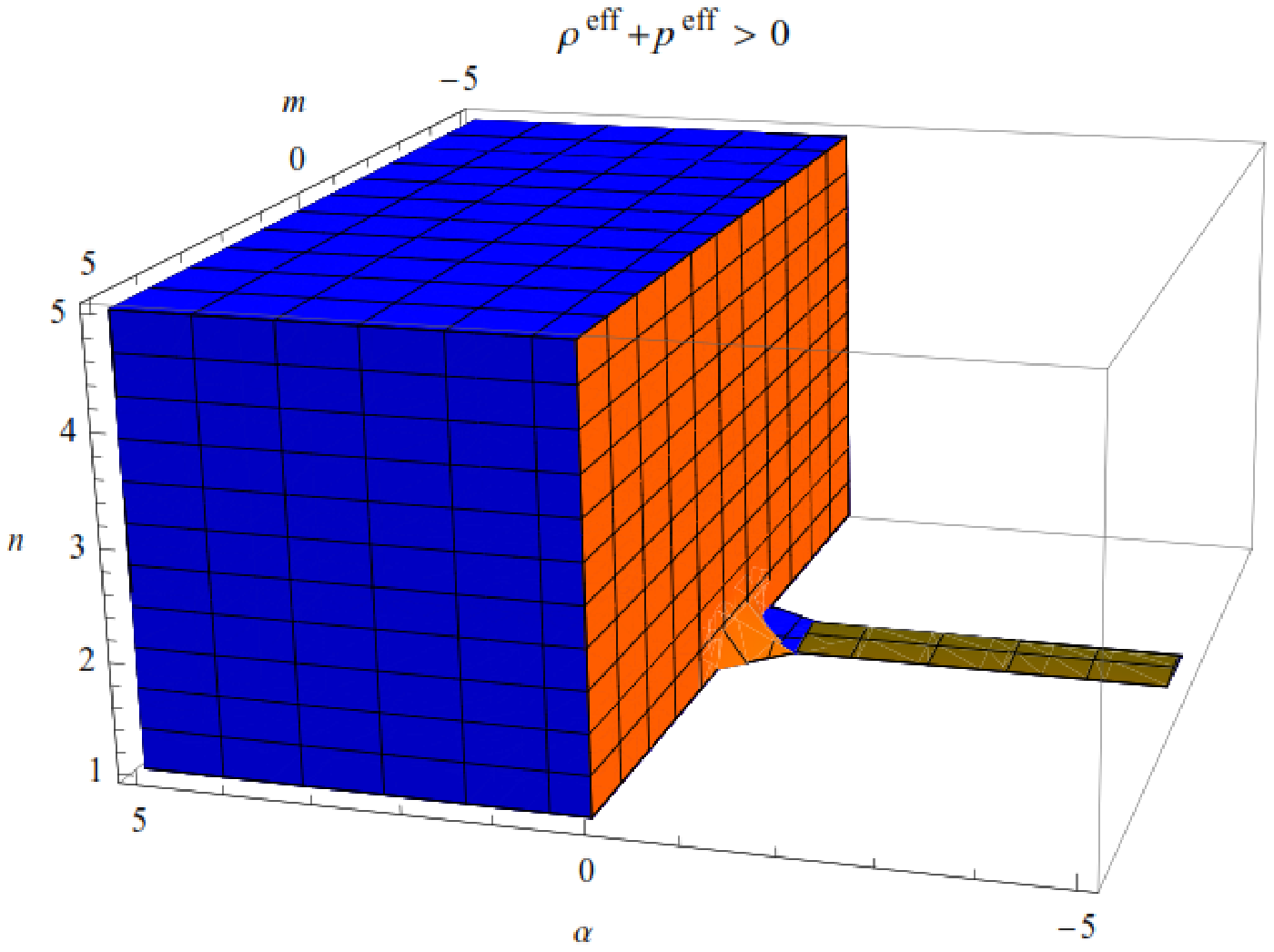,width=.48\linewidth} \caption{WEC validity regions
for $f(G)$ model mentioned in Eq.(\ref{model2}), Here left plot shows
$\rho^{\textrm{eff}}>0$ while the right plot indicates
$\rho^{\textrm{eff}}+ p^{\textrm{eff}}>0$ with respect to $\alpha$,
$m$ and $n$ with $\beta=1$.} \label{2f2}
\end{figure}

Next, we consider another realistic formulation of $f(G)$ model
\cite{16}
\begin{equation}\label{model2}
f(G)= \alpha  G^n \left(\beta  G^m+1\right),
\end{equation}
where $\alpha,~\beta$ and $m$ are the arbitrary constants and $n$ is
a positive constant. This model could be useful to understand the
finite time future singularities \cite{no08}. The outcomes of this
model are found to be in agreement with
the local test as well as the cosmological bounds \cite{no07}.\\
By making use of Eq.(\ref{model2}), the effective energy density has
been found to be
\begin{align}\nonumber
{\rho ^{\textrm{eff}}}&= {24^n}\alpha {\left( { - {H^4}q}
\right)^n}( - 1 + n - {24^m} \beta {\left( { - {H^4}q} \right)^m} +
{24^m}m\beta {\left( { - {H^4}q} \right)^m}+ {24^m}n\beta\\\nonumber
 &\times {\left( { - {H^4}q} \right)^m} - \frac{1}{{{q^2}}}
 \left( {j + q\left( {3 + 2q} \right)} \right)({24^m}( - 1 + m)m\beta {\left( { - {H^4}q}
 \right)^m}+ {n^2}\{1 +\\\label{2ro}
 &\times{{24}^m}\beta{{\left( { - {H^4}q} \right)}^m}\}+
 n\left( { - 1 + {{24}^m}\left( { - 1 + 2m} \right)\beta {{\left( { - {H^4}q} \right)}^m}} \right)))\geq 0,
\end{align}
while the combination of effective pressure and energy density
becomes
\begin{align}\nonumber
&{\rho ^{\textrm{eff}}} + {p^{\textrm{eff}}}= \frac{1}{{{q^3}}}{3^{
- 1 + n}}{8^n}\alpha {\left( { - {H^4}q} \right)^n}( - 3q\left( {j +
q\left( {3 + 2q} \right)} \right)({24^m}( - 1 + m)m\beta {\left( { -
{H^4}q} \right)^m}\\\nonumber &+ {n^2}\left( {1 + {{24}^m}\beta
{{\left( { - {H^4}q} \right)}^m}} \right) + n( - 1 + {24^m}( - 1 +
2m)\beta {\left( { - {H^4}q} \right)^m}))-(j + q(3 \\\nonumber &+
2q))^2({24^m}m \left( {2 - 3m + {m^2}} \right)\beta {\left( { -
{H^4}q} \right)^m} +{n^3}\left( {1 + {{24}^m}\beta {{\left( { -
{H^4}q} \right)}^m}} \right)+ 3{n^2}\\\nonumber &(-1+{{24}^m}(- 1 +
m)\beta {{\left( { -
 {H^4}q} \right)}^m}) + n\left( {2 + {{24}^m}\left( {2 - 6m + 3{m^2}}
 \right)\beta {{\left( { - {H^4}q} \right)}^m}} \right))\\\nonumber
&- q({24^m}\left( { - 1 + m} \right)m\beta {\left( { - {H^4}q}
\right)^m} +{n^2}\left( {1 + {{24}^m}\beta {{\left( { - {H^4}q}
\right)}^m}} \right) + n( - 1 + {24^m}
\\\label{2ropp}
&\times(-1+2m)\beta{\left( { - {H^4}q} \right)^m}))\left( { - 3 + 5q
+ 18{q^2} + 6{q^3} + j\left( {4 + 8q} \right) - s} \right))\geq 0.
\end{align}
These two inequalities (\ref{2ro}) and (\ref{2ropp}) are much
complicated to find the exact analytical expression for the
parameters. So, we shall fix some parameters by putting them equal to
specific values. For simplicity, we let $\alpha=1,\beta=1$ and plot
$\rho^{\textrm{eff}}$ and $\rho^{\textrm{eff}}+ p^{\textrm{eff}}$
which are the function of $m$ and $n$ only, as shown in fig.\ref{2f1}.
It can be observed from this figure that WEC is also valid for the model (\ref{model2}).\\
Now, we will check the constraints on parameters for the validity of
WEC. For this purpose, let $\beta=1$ and we found those regions
under which WEC is valid. For $\rho^{\textrm{eff}}\geq 0$, we require\\
(1) $\alpha>0$ with any value of $m$.\\
(2) $m<0$ with a very small $n$.\\
Similarly for $\rho^{\textrm{eff}}+ p^{\textrm{eff}}\geq 0$, we require\\
(1) $\alpha>0$ with any value of $m$.\\
(2) $-1<m<0$ with a very small $n$.\\
The region plots for WEC are shown in fig.\ref{2f2}, in which the
left plot is for the effective energy density while the right plot
is for the summation of effective energy density and pressure.

\subsection{Model 3}

\begin{figure} \centering
\epsfig{file=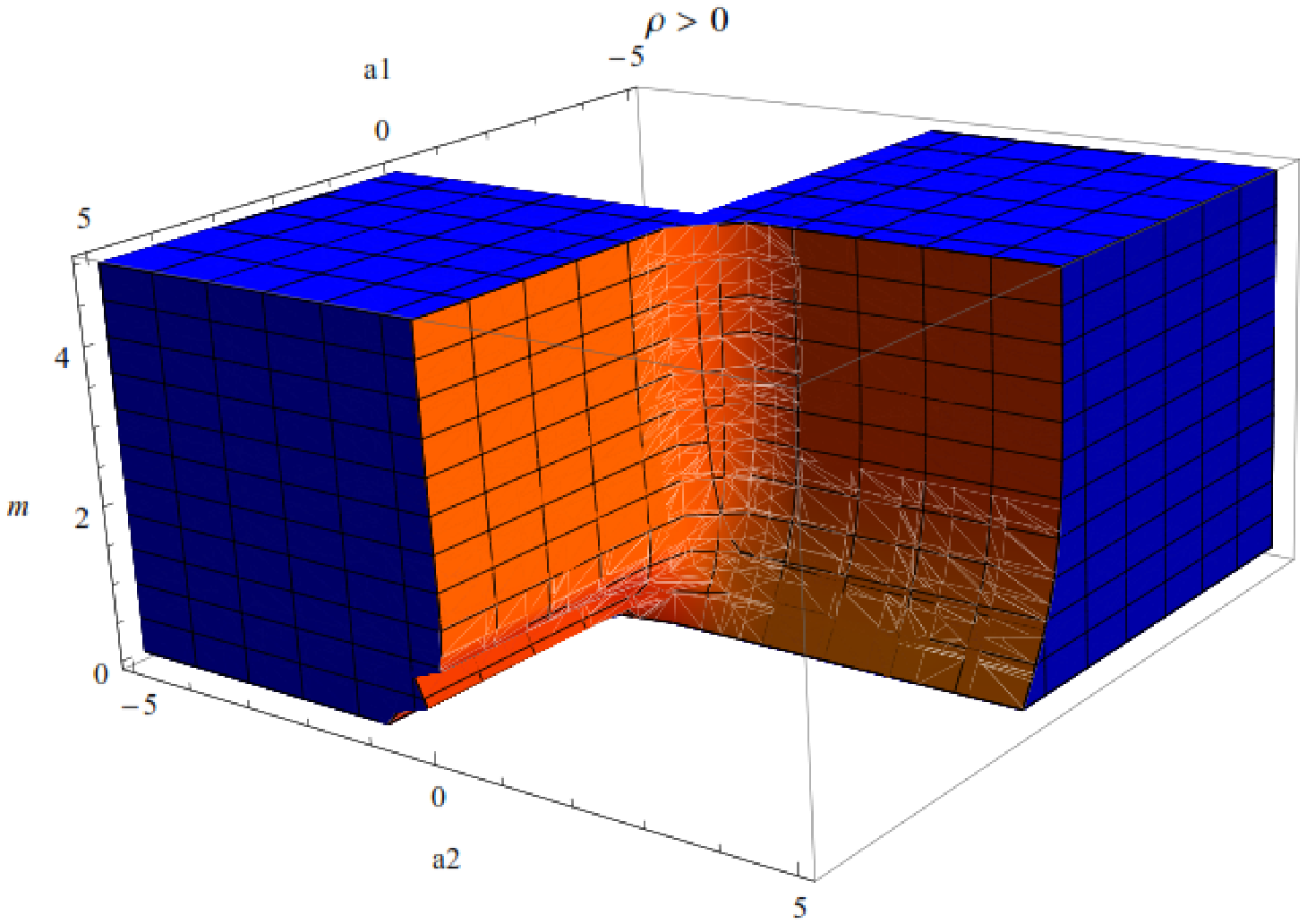,width=.48\linewidth}
\epsfig{file=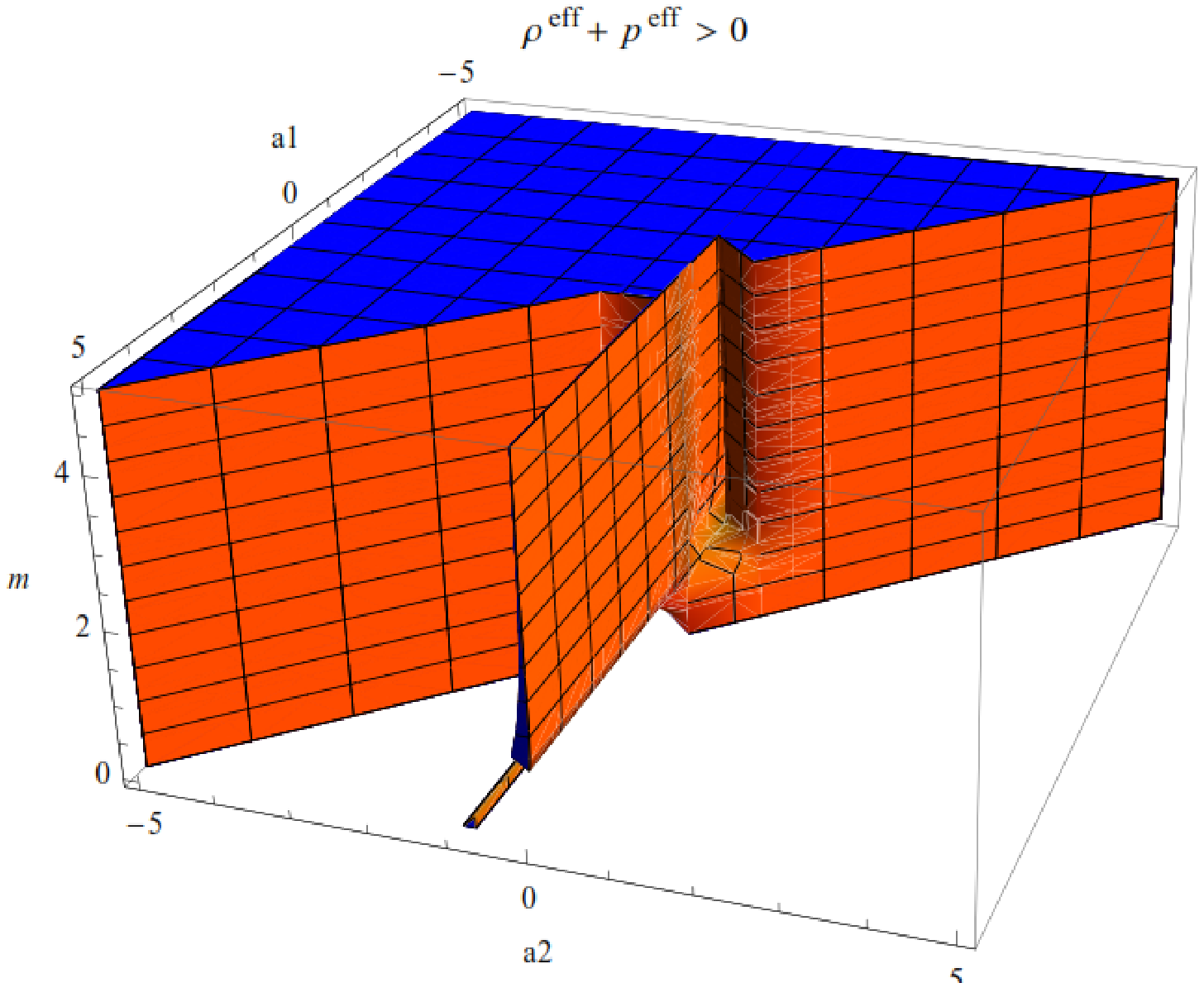,width=.48\linewidth} \caption{WEC validity epochs
for $f(G)$ model given in Eq.(\ref{model3}). Here, left plot indicates $\rho_{eff}>0$,
while the right plot describes $\rho^{\textrm{eff}}+
p^{\textrm{eff}}>0$.} \label{3f1}
\end{figure}

It would be interesting to analyze another realistic model in $f(G)$
gravity \cite{15}
\begin{equation}\label{model3}
f(G) = \frac{{{a_1} {G^n} + {b_1}}}{{{a_2} {G^n} + {b_2}}},
\end{equation}
where $a_1,~b_1,~a_2,~b_2$ and $n$ are the arbitrary constants, with
$n>0$. This model could be helpful in the study of finite time
future singularities as well as the late time cosmic acceleration.
The effective energy density for this model becomes
\begin{align}\nonumber
{\rho ^{eff}}&=\frac{{ - 1}}{{{{\left( {{a_2}{{24}^m}{{\left(
{-{H^4}q} \right)}^m} + {b_2}} \right)}^3}}}
(\frac{{{{24}^m}m}}{{{q^2}}}\left( {j + q\left( {2q + 3} \right)}
\right)({a_2}{b_1}\\\nonumber
 &- {a_1}{b_2}){\left( { - H{{\left( t \right)}^4}q} \right)^m}\left( {{a_2}{{24}^m}\left( {m + 1}
 \right){{\left( { - {H^4}q} \right)}^m} - {b_2}m + {b_2}}
 \right)\\\nonumber
 &+ {24^m}m\left( {{a_2}{b_1} - {a_1}{b_2}} \right){\left( { - {H^4}q} \right)^m}\left(
 {{a_2}{{24}^m}{{\left( { - {H^4}q} \right)}^m} + {b_2}}
 \right)\\\label{3ro}
& + \left( {{a_1}{{24}^m}{{\left( { - {H^4}q} \right)}^m} + {b_1}}
\right)
 {\left( {{a_2}{{24}^m}{{\left( { - {H^4}q} \right)}^m} + {b_2}} \right)^2})\geq 0,
\end{align}
and the combination of the effective energy density and pressure
becomes
\begin{align}\nonumber
{\rho ^{eff}} + {p^{eff}} &= \frac{1}{{{q^3}{{\left(
{{a_2}{{24}^m}{{\left( { - {H^4}q} \right)}^m} + b2}
\right)}^4}}}{3^{m - 1}}{8^m}m({a_2}{b_1}- {a_1}{b_2}){\left( { -
{H^4}q} \right)^m}\\\nonumber & \times({\left( {j + q\left( {2q + 3}
\right)}
 \right)^2}({a_2}^2{576^m}({m^2}+ 3m + 2){\left( { - {H^4}q} \right)^{2m}} + {a_2}{b_2}\\\nonumber
&\times({ - {2^{3m + 2}}}){3^m}\left( {{m^2} - 1} \right){\left( { -
{H^4}q}
 \right)^m}+ {b_2}^2\left( {{m^2} - 3m + 2} \right))\\\nonumber
 & - q(j\left( {8q + 4} \right) + 6{q^3} + 18{q^2} +
 5q- s - 3)({a_2}{{24}^m}({ - {H^4}q})^m\\\nonumber
 &+ {b_2})({a_2}{24^m}\left( {m + 1} \right){\left( { - {H^4}q}
 \right)^m}- {b_2}m + {b_2}) - 3q\left( {j + q\left( {2q + 3} \right)}
\right)\\\label{3ropp} & \times ({a_2}{24^m}(- {H^4}q)^m+
{b_2})\left( {{a_2}{{24}^m}\left( {m + 1} \right){{\left( { -
{H^4}q} \right)}^m} - {b_2}m + {b_2}} \right))\geq 0.
\end{align}
As this model contains five parameters, i.e., $a_1,~a_2,~b_1,~b_2,~m$,
so we will fix some of these parameters by assigning some specific values.
For simplicity, we let $b_1=-1, b_2=1$. Now, the constraints on other
parameters are (for $\rho_{eff}\geq 0$ with $m>0$)\\
(1) $a_1<0$ with $a_2>-1$.\\
(2) $a_1>1$ with $a_2<0$.\\
One can can easily check above mentioned constraints through the
left plot of fig.\ref{3f1}. The validity regions for
$\rho^{\textrm{eff}}+ p^{\textrm{eff}}\geq 0$ would impose some
constraints on the parameters $a_1$ and $a_2$. It is seen that $a_1$
depends on the choice of $a_2$ but if $a_1<0$ and $a_2<0$ with $m>0$
then these give $\rho^{\textrm{eff}}+ p^{\textrm{eff}}\geq 0$ as
shown in the right plot of fig.\ref{3f1}.

\section{Summary}

\vspace{0.5cm}

In the present paper, we have explored the influence of modified GB
gravity models on the existence of realistic configurations of cosmological
perfect fluid models. The investigation of ECs are closely
associated with the realistic picture of the traversable WH
solutions. To avoid the use of exotic matter content at the WH
throat, the exploration of viable and well-consistent models is an
alluring objective. We have considered the behavior of FLRW metric
filled with an ideal fluid. The $f(G)$ field equations turn out to
be highly non-linear that could not be solved without taking certain
physically consistent assumptions. From $f(G)$ field equations, we
have evaluated general energy inequalities relation. We have considered
three different modified GB gravity models, i.e., $f(G)=\alpha {G^n}
+ \beta G\log[G],~f(G)= \alpha  G^n \left(\beta G^m+1\right)$ and
$f(G) = \frac{{{a_1} {G^n} + {b_1}}}{{{a_2} {G^n} + {b_2}}}$. We
have checked the behavior of ECs by taking into account all of the
above mentioned modified GB models and perfect fluid. Then, the
recent calculated values of the parameter Hubble, deceleration, jerk
and snap are used with the different specific viable $f(G)$ models.
We plotted the regions where NEC and WEC hold against various
parameters of $f(G)$ gravity. The graphical features show some
results given as follows:\\\\
(i) By considering higher curvature corrections induce from
$f(G)=\alpha {G^n} + \beta G\log[G]$ model, the effectiveness of WEC
could be attained by setting positive values of $\alpha$ and $n>1$
with any $\beta\in(-5,5)$ or by taking $\alpha$ and $\beta$ to be
positive with $n$ less than $-1$. Further, we claimed that the
breaching of WEC could be nullified by considering positive values of
($n,~\alpha,~\beta$) tetrad. The validity of NEC could be achieved
by taking $n>1$ along with the positive values of $\alpha$ for any
$\beta$ or by setting $\beta$ to
be positive and very little value of $n$ with for all $\alpha$.\\\\
(ii) In the realm of $f(G)= \alpha  G^n \left(\beta G^m+1\right)$,
the WEC would be valid under two possibilities. One with positive
$\alpha$ with any $m$ and other with negative $m$ and very little
value of $n$. The validity of NEC could be attained by setting
$m\in(-1,0)$ with small $n$ or by considering positive numeric value
of $\alpha$ with any $m$. It is worthy to mention that in this
analysis, we have assumed $\beta$ to be unity.\\\\
(iii) Our next considered model is quiet complicated, as it
comprises of five parameters, i.e., $a_1,a_2,b_1,b_2,m$. In order to
handle such situation, we have fixed some of these parameters to get
estimated validity epochs of WEC and NECs. Thus, in $f(G) =
\frac{{{a_1} {G^n} + {b_1}}}{{{a_2} {G^n} + {b_2}}}$, the WEC will
be valid, if one takes $b_1=-1,~b_2=1$ along with negative value of
$a_1$, positive value of $m$ and $a_2$ to be greater than -1.
Further, by setting positive $m$, negative $a_2$ and $a_1$ to be
greater than unity (for details see fig.\ref{3f1}). The NEC
violation could be avoided by taking negative $a_1$ and $a_2$ with
$m>0$ as shown in fig.\ref{3f1}.

Finally, it is remarked that the exploration of viable $f(G)$ models
performed in this paper could easily be extended for the case that
there exists convenient usual complicated matter content within FLRW
metric. The corresponding analysis may lead to some significant
qualitative outcomes in comparison with the discussion of pure
gravity. It will be executed elsewhere.

%%%%%%%%%%%%%%%%%%%%%%%%
%%%  Acknowledgments
%%%%%%%%%%%%%%%%%%%%%%%%
\section*{Acknowledgments}

This work was partially supported by the JSPS KAKENHI Grant Number
JP 25800136 and the research-funds presented by Fukushima University
(K.B.).

\end{document}